\documentclass[pdflatex,sn-mathphys-num]{sn-jnl}

\usepackage{graphicx}%
\usepackage{multirow}%
\usepackage{amsmath,amssymb,amsfonts}%
\usepackage{amsthm}%
\usepackage{mathrsfs}%
\usepackage[title]{appendix}%
\usepackage{xcolor}%
\usepackage{textcomp}%
\usepackage{manyfoot}%
\usepackage{booktabs}%
\usepackage{algorithm}%
\usepackage{algorithmicx}%
\usepackage{algpseudocode}%
\usepackage{listings}%
\usepackage[T1]{fontenc}

\theoremstyle{thmstyleone}%
\theoremstyle{thmstyletwo}%

\theoremstyle{thmstylethree}%

\begin{document}

\title[Article Title]{All-Optical Wide-Field Magnetometry with Van Der Waals Quantum Sensor}


\author*[1]{\fnm{Feifei} \sur{Zhou}}\email{ffzhou@cjlu.edu.cn}

\author[2]{\fnm{Peiyan} \sur{Ma}}

\author[1]{\fnm{Jiajun} \sur{Li}}

\author[1]{\fnm{Ke} \sur{Jing}}

\author[3]{\fnm{Shihao} \sur{Ru}}

\author[1]{\fnm{Hongwei} \sur{Chen}}

\author*[1]{\fnm{Ying} \sur{Dong}}\email{yingdong@cjlu.edu.cn}

\author*[1]{\fnm{Xinqing} \sur{Wang}}\email{wxqnano@cjlu.edu.cn}

\affil[1]{\orgdiv{College of Metrology Measurement and Instrument}, \orgname{China Jiliang University}, \orgaddress{\city{Hangzhou 310018}, \country{China}}}

\affil[2]{\orgdiv{Department of Physics}, \orgname{Southern University of Science and Technology}, \orgaddress{\city{Shenzhen 518055}, \country{China}}}

\affil[3]{\orgdiv{School of Electrical and Electronic Engineering}, \orgname{Nanyang Technological University}, \orgaddress{\city{Singapore 639798}, \country{Singapore}}}


\abstract{Negatively charged boron vacancy ($V_B^-$) centers in hexagonal boron nitride ($h$-BN) have attracted wide-range interests owing to their van der Waals lattice and their potentials for $in$-$situ$ quantum sensing. Here we propose and experimentally demonstrate an all-optical strategy for wide-field magnetometry based on $V_B^-$ centers. This strategy exploits the magnetically sensitive ground-state level anti-crossing (GSLAC) of $V_B^-$ centers, which induces a strong electron spin transition between $m_S = 0$ and $m_S = -1$ states, enabling microwave-free magnetic field measurement. By monitoring the shift of GSLAC feature, the external magnetic field can be precisely determined. Using this technique, we demonstrate all-optical wide-field imaging of near-field DC magnetic field distribution from current-carrying circuits over an area of around 42 $\times$ 21 $\mu$m$^2$. An estimated photon shot-noise-limited sensitivity of 67.1 $\mu$T/$\sqrt{\text{Hz}}$ is achieved for a single pixel, which is an approximately threefold improvement over the ODMR method, along with a spatial resolution of about 1 $\mu$m per pixel. Our approach expands the applicability of $V_B^-$ centers in quantum sensing, paving the way for robust and convenient magnetometry under extreme conditions.}




\maketitle

\section{Introduction}\label{sec1}
Optically addressable spin defects in wide band-gap van der Waals (vdW) materials have emerged as promising platforms for the quantum sensing of various physical quantities. The mostly studied two-dimensional spin system is the negatively charged boron vacancy ($V_B^-$) center in hexagonal boron nitride ($h$-BN), which has been employed for the detection of magnetic fields\cite{magnetic1, magnetic2, magnetic3, magnetic4}, temperature\cite{temperature1, temperature2, temperature3}, strain\cite{strain1, strain2, strain3}, and paramagnetic spins\cite{paraspin1, paraspin2, paraspin3, paraspin4}. $V_B^-$ centers offer distinct advantages for $in$-$situ$ quantum sensing, as they can be fabricated into few-layer films and transferred directly onto the target surface very closely. By exploiting measurement protocols such as optically detected magnetic resonance (ODMR) and dynamical decoupling and coherent dynamics, $V_B^-$ centers enable the precise detection of both DC and alternating magnetic fields. In these applications, electron spin resonance (ESR) serves as the fundamental mechanism underlying the sensing process.

Conventional ESR measurements of $V_B^-$ centers rely on resonant microwave (MW) fields to probe shifts in spin energy levels through MW frequency sweeping, while this approach faces limitations in certain experimental contexts. First, it significantly introduces the complexity of measurement systems, which requires high-speed synchronization devices, MW sources and amplifiers, as well as high-efficiency MW antennas. Second, since the coherence time of $V_B^-$ centers is quite short\cite{coherence}, a strong MW power is desirable for achieving a good signal-to-noise ratio of ESR, while it simultaneously generates substantial heating effects that compromise the measurement accuracy. This issue is especially critical for quantum sensing tasks performed under extreme conditions, such as cryogenic temperatures or high pressures. Therefore, developing MW-free magnetometry methods is essential for expanding the practical applications of $V_B^-$ centers.

In this work, we propose and experimentally demonstrate an all-optical magnetometry strategy with $V_B^-$ centers. This strategy exploits the ESR of $V_B^-$ centers occurring near the ground-state level anti-crossing (GSLAC), where the $m_S = 0$ and $m_S = -1$ states become degenerate, thereby inducing a strong ESR transition without the assistance of resonant MW fields. We first investigate the magnetic-field-dependent ODMR behaviors of $V_B^-$ centers to identify the GSLAC feature. Since the GSLAC enables a spontaneous ESR transition that results in the disappearance of ODMR signal, we then measure the magnetic-field-dependent photoluminescence (PL) intensity of $V_B^-$ centers to precisely confirm the central field of GSLAC feature. By applying another magnetic field to be measured, we observe an obvious shift in the central field and verify the feasibility of the GSLAC-based magnetometry strategy. Combining with spin-based wide-field measurement, we realize the wide-field microscopy of direct-current (DC) magnetic fields genetrated by a current-carrying circuit over an area of approximately 42 $\times$ 21 $\mu$m$^2$. We also quantitatively study the performance of our GSLAC strategy by varying the current strengths. Finally, we achieve a photon shot-noise-limited sensitivity of approximately 67.1 $\mu$T/$\sqrt{\text{Hz}}$ for a single pixel and an estimated spatial resolution of about 1 $\mu$m. Our work provides a promising all-optical and $in$-$situ$ magnetometry strategy based on GSLAC mechanism of $V_B^-$ centers, which significantly reduces the hardware threshold and is beneficial for the development of integrated quantum sensing platform. Moreover, our method can be extended to the wide-range studies of biology, material science, and condensed matter physics even under extreme conditions, such as cryogenic temperatures and high pressures.

\section{Methods}\label{sec2}
\begin{figure*}[htbp]
  \centering
  \includegraphics[width=1\textwidth]{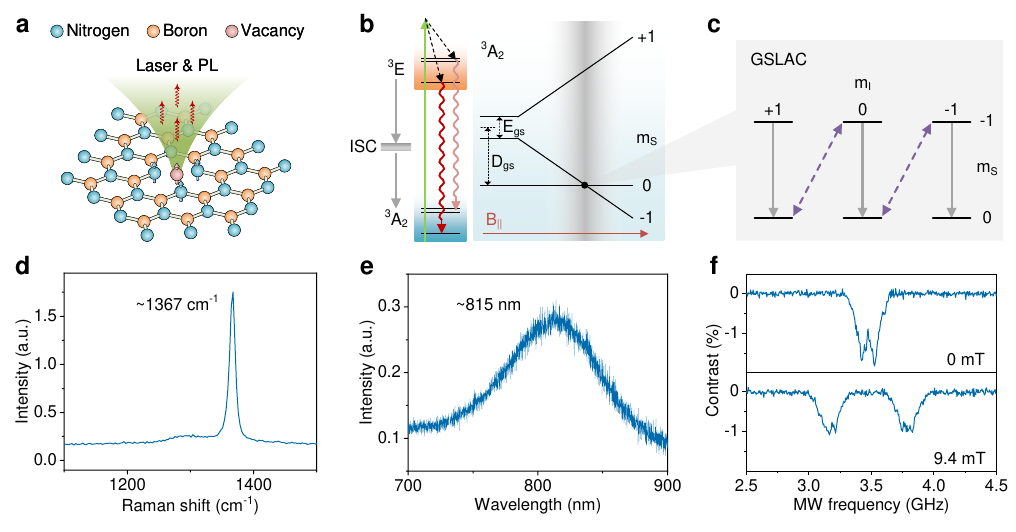} 
  \caption{(a) Atomic lattice structure of the $V_B^-$ center in $h$-BN. (b) Schematic of electron-spin energy levels of the $V_B^-$ center. Green arrow represents the 532-nm laser irradiation, red wave arrows denotes the photon emissions, and the gray lines indicate the ISC transition of the $V_B^-$ center. Orange (blue) gradient-filled block shows the excited (ground) triplet spin states. The black dot represents the GSLAC point. (c) Schematic of spin transitions at GSLAC point. The gray arrows denote the ISC transition, and the purple dashed arrows denote the ESR assisted by the flip-flop of nuclear spins. (d) Raman spectrum of the $h$-BN sample used in this study. An expected peak of around 1367 cm$^{-1}$ is observed. (e) Room-temperature PL spectrum of $V_B^-$ centers in $h$-BN flake. The central PL wavelength is about 815 nm. (f) ODMR spectra of $V_B^-$ centers with (without) an external magnetic field of 9.4 mT. A small splitting of $m_s = \pm 1$ in the upper spectrum is attributed to the intrinsic strain coupling.}
  \label{fig1}
\end{figure*}

The $V_B^-$ center is a spin-1 electron spin system, consisting of a boron vacancy and nearby three nitrogen nuclei, as shown in Fig.~\ref{fig1}(a). In the presence of an aligned bias magnetic field $B_\parallel$, the ground-state Hamiltonian of the $V_B^-$ center can be described as
\begin{equation}
H_{gs} = D_{gs} S_z^2 + E_{gs} (S_x^2 - S_y^2) + \gamma_e B_\parallel S_z + \sum_{i=1}^3 A_{zz} I_z^i S_z,
\end{equation}
where $D_{gs}$ ($E_{gs}$) is the zero-field splitting parameter induced by spin-spin (spin-strain) couplings, $S_{x,y,z}$ and $I_z^i$ denote the electron-spin and the $i$-th nuclear-spin operators, $\gamma_e$ is the gyromagnetic ratio of electron spin, and $A_{zz}$ represents the hyperfine coupling strength of nitrogen nucleus. Similar to the nitrogen-vacancy center in diamond, the $V_B^-$ center also exhibits a triplet energy-level structure, as indicated in Fig.~\ref{fig1}(b). With the illumination of a 532 nm laser, the electron spins are pumped from ground triplet states ($^3\text{A}_2$) to excited triplet states ($^3$E) and then spontaneously decay back to the $^3\text{A}_2$ states and emits the wide-band fluorescence. In this process, a small population of electron spins at $m_s = \pm 1$ states decays back to the $m_s = 0$ through an intersystem crossing (ISC) transition without emission, resulting in a spin-state-dependent fluorescence intensity and enabling the optical readout of the electron spin state.

In the $^3\text{A}_2$ states, the $m_s = \pm 1$ states can be further split by the bias field $B_z$ due to the Zeeman effect. With increasing $B_z$, the $m_s = -1$ state continuously approaches and eventually reaches the $m_s = 0$ state. At this point, the wave functions of the electron spin and nuclear spins become mixed, enabling a strong interaction which exchanges the spin angular momentum among these spins\cite{gslac}. Assisted by this interaction, the population of electron spins is transferred equally to both $m_s = -1$ and $m_s = 0$ states. This process is called GSLAC, which allows the MW-free ESR transition of the $V_B^-$ centers, as illustrated in Fig.~\ref{fig1}(c).

To perform this study, we exfoliate the bulk $h$-BN into thin flakes with a thickness of
around 1 $\mu$m, transfer the $h$-BN flake onto a gold strip-line with a width of 20 $\mu$m, and then bombard the sample using proton beams with a 250 keV energy and a dose of 3 $\times$ 10$^{16}$ cm$^{-2}$ to create $V_B^-$ centers. To confirm the generation, we then measure Raman and PL spectra of our sample using a commercial spectrometer (Horiba XploRA Plus), and the results are presented in Fig.~\ref{fig1}(d-e). Both Raman characteristics of the $h$-BN flake and PL features of the $V_B^-$ centers are observed. We also measure the ODMR spectra of our sample with (without) an external magnetic field of about 9.4 mT to confirm the generation of $V_B^-$ centers. All results agree well with expectations. According to the results shown in Fig.~\ref{fig1}(f), the parameters $D_{gs}/h$ and $E_{gs}/h$ of our sample are around 3.48 GHz and 50 MHz, respectively, which agree well with previous reports\cite{zerosplitting1, zerosplitting2}.

\begin{figure*}[htbp]
  \centering
  \includegraphics[width=1\textwidth]{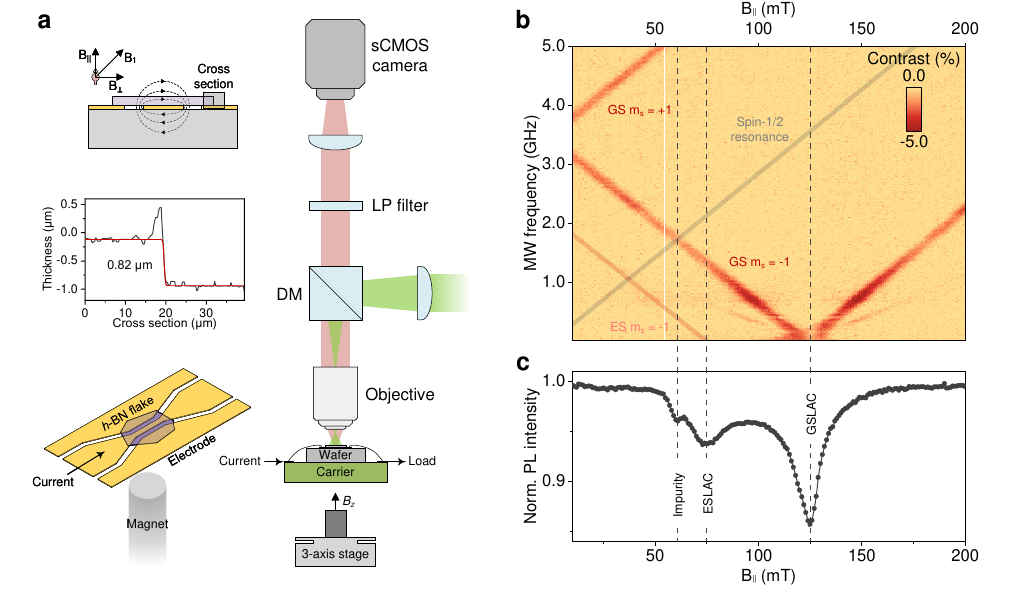}
  \caption{(a) The architecture of our wide-field microscope. Green-filled path represents the 532 nm laser, and the light red-filled path denotes the collection of PL signals. The three-dimensional model exhibits the core details of CPW circuits covered by $h$-BN flake with a thickness of 0.82 $\mu$m. A cross section schematic of CPW describes the contour lines of current-induced magnetic field $B_1$, which consists of two components $B_\parallel$ and $B_\perp$ referring to the quantum axis of $V_B^-$ centers. DM, dichroic mirror; LP, long-pass. (b) Magnetic-field-dependent ODMR mapping of $V_B^-$ centers. The dark red tracks are attributed to the ground-state ESR between $m_s = 0$ and $m_s = \pm 1$. The other two transparent tracks are manually marked to point out the excited-state ESR between $m_s = 0$ and $m_s = -1$ (red) and couplings between $V_B^-$ centers and spin-1/2 impurities in $h$-BN (gray). (c) Magnetic-field-dependent PL intensities of $V_B^-$ centers. Three dashed lines are marked between (b) and (c) for clear comparison.}
  \label{fig2}
\end{figure*}

To demonstrate our strategy, we build a spin-based wide-field microscope shown in Fig.~\ref{fig2}(a). A continuous-wave 532 nm laser is reflected in an objective (Zeiss 50$\times$/0.75) and focused at its rear focal point to realize a large area of illumination. Fluorescence is collected, filtered and focused in a scientific Complementary Metal-Oxide-Semiconductor (sCMOS) camera (Thorlabs CS2100M-USB) to realize the fluorescence mapping of $V_B^-$ centers. A golden coplanar waveguide (CPW) is fabricated on a silicon wafer to carry the current, which is covered by our $h$-BN flake and connects to a carrier by bonding aluminum wires. Using a commercial confocal microscope (CHOTEST VT6100), the thickness of our $h$-BN film is measured to be approximately 0.82 $\mu$m. A cylindrical permanent magnet is mounted on a 3-axis stage to provide a varying magnetic field. Since the magnetic field strength attenuates nonlinearly with increasing distance, this relationship is pre-calibrated to ensure a linear increase in the magnetic field strength.

\section{Results and analysis}\label{sec3}
To investigate the GSLAC features, we continuously vary the magnetic field strength $B_\parallel$ from 10 mT to 200 mT and conduct the ODMR measurements simultaneously. For a single ODMR measurement, the camera acquires two signals, $S_1$ with MW and $S_2$ without MW. The ODMR contrast is then calculated by $(S_1 - S_2) / (S_1 + S_2)$. The ODMR behaviors with increasing $B_\parallel$ are presented in Fig.~\ref{fig2}(b), where two dark-red stripes exhibit the ESR transitions of $m_s = 0 \leftrightarrow m_s = +1$ (upper) and $m_s = 0 \leftrightarrow m_s = -1$ (lower), respectively. The lower stripe approaches near zero and then rises, and the ODMR signal at the turning point disappears due to the GSLAC mechanism. Two transparent tracks are manually marked in the ODMR mapping. The red track describes the expected excited-state ODMR signals, which is difficult to observe due to the room-temperature noise. The gray track describes the potential ESR signal probably coming from spin-1/2 impurities in $h$-BN\cite{impurities}. To further locate the GSLAC point, we also monitor the magnetic-dependent fluorescence of $V_B^-$ centers by strengthening the $B_\parallel$ from 10 mT to 200 mT and acquire the PL intensity using camera simultaneously. The results involving three dips are illustrated in Fig.~\ref{fig2}(c). The first dip at around 60 mT might be caused by the strong coupling between $V_B^-$ centers and the above-mentioned impurities. The other two dips occurring at about 75 mT and 125 mT are attributed to excited-state level anti-crossing (ESLAC) and GSLAC interactions of $V_B^-$ centers\cite{dips}, respectively. According to three marked dashed lines, this magnetic-dependent fluorescence trend agrees well with the ODMR mapping in Fig.~\ref{fig2}(b).

\begin{figure*}[htbp]
  \centering
  \includegraphics[width=1\textwidth]{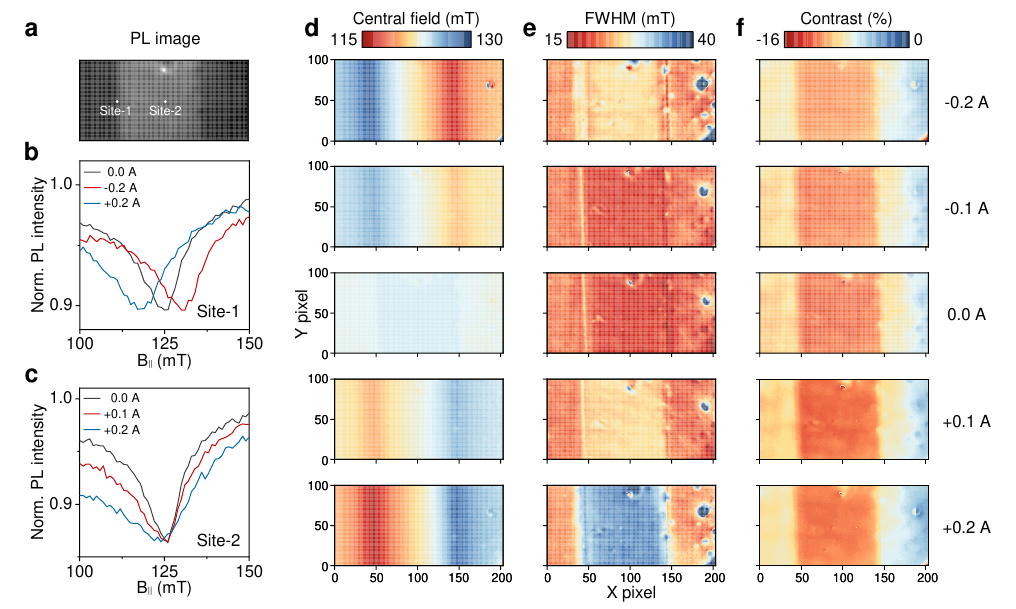}
  \caption{(a) PL image of $h$-BN flake covering the strip-line surface. GSLAC behaviors at site-1 and site-2 are further studied. (b) GSLAC behaviors at site-1 with current magnitudes of 0.0 A and $\pm$0.2 A. Significant shifts of central field are observed. (c) GSLAC behaviors at site-2 with current magnitudes of 0.0 A, +0.1 A, and +0.2 A. The central field has no shifts while the FWHM is significantly broadened due to the increasing of $B_\perp$. (d-f) Distribution of GSLAC features including the central fields (d), the FWHMs (e), and the contrasts (f).}
  \label{fig3}
\end{figure*}

We then study the magnetic-field-induced shift of GSLAC point. A DC current is produced by a commercial current source (Keithley 2231A-30-3) and delivered to the golden CPW. According to Amp\`ere's law, this current-carrying circuit can generate a surrounding DC magnetic field. We choose an area of 42 $\times$ 21 $\mu$m$^2$ on the strip-line surface of CPW. Fig.~\ref{fig3}(a) presents the PL image of $h$-BN flake covering this area. We first focus on the site-1 pixel and measure the GSLAC features with different currents (0.0 A, -0.2 A, +0.2 A) for comparison. The results are presented in Fig.~\ref{fig3}(b). Without the DC current, the central field of GSLAC is about 125 mT. With a current of +0.2 A (-0.2 A), the central field shifts to the left (right), which indicates that applying a current of +0.2 A generates a DC field aligned with (opposite to) the bias magnetic field. We then focus on the site-2 pixel and conduct the GSLAC measurement with currents of 0.0 A, +0.1 A, and +0.2 A. The results are shown in Fig.~\ref{fig3}(c). At this time, the central field of GSLAC has no obvious shifts while the full width at half maximum (FWHM) of GSLAC dip are broadened with increasing current. This is expected, as the current-induced magnetic field at the site-2 pixel is nearly perpendicular to the quantum axis of the $V_B^-$ centers, thereby enabling the detection of the perpendicular magnetic field component $B_\perp$. These results confirm the feasibility and performance of our scheme.

We further extend the GSLAC magnetometry to all pixels in this 42$\times$21 $\mu$m$^2$ area and realize the magnetic field imaging of current-carrying strip-line. We conduct the same measurement of magnetic-dependent PL intensity using camera and extract the GSLAC features of $V_B^-$ centers in all pixels. Here we conduct five groups of measurements under different currents (-0.2 A, -0.1 A, 0.0 A, 0.1 A, 0.2 A). We then use the Lorentzian function to fit the contrasts, FWHMs, and central fields of GSLAC features for all pixels, and these results are performed in the heatmaps of Fig.~\ref{fig3}(d-f).
In the gap area between strip-line and ground-line, an expected gradient of central fields is observed, as illustrated in Fig.~\ref{fig3}(d). This gradient becomes more pronounced with increasing current. Moreover, reversing the current direction also reverses the gradient. Note that the central field is mainly dominated by the longitudinal component $B_{\parallel}$ (parallel to the quantum axis of $V_B^-$ centers), which explains why the central fields on the center of strip-line surface remain nearly unchanged. Since the transverse component $B_\perp$ can broaden the FWHM of GSLAC dip, we also conduct the FWHM mapping to evaluate the distribution of $B_\perp$ with different currents. The results are presented in Fig.~\ref{fig3}(e). The current-induced magnetic field on the strip-line surface is oriented perpendicular to the quantum axis of $V_B^-$ centers, leading to large FWHM values on the strip-line surface. For the area off the surface, no obvious $B_{\perp}$ is observed, consistent with expectations. Notably, the FWHM distributions at 0.0 A and -0.1 A ( as well as at +0.1 A and -0.2 A) remain symmetric. This phenomenon is attributed to a residual $B_{\perp}$ resulting from a slight mismatch between bias magnet field and the quantum axis of $V_B^-$ centers. Furthermore, we study the contrast behaviors of GSLAC dip under different currents and perform the mapping results in Fig.~\ref{fig3}(f). The contrast is nearly independent of the current magnitude. The contrast values on the strip-line surface are larger that those off the surface, owing to the improved fluorescence collection efficiency and plasmonic enhancement from the gold substrate\cite{plasmonic1, plasmonic2}.

\begin{figure*}[htbp]
  \centering
  \includegraphics[width=1\textwidth]{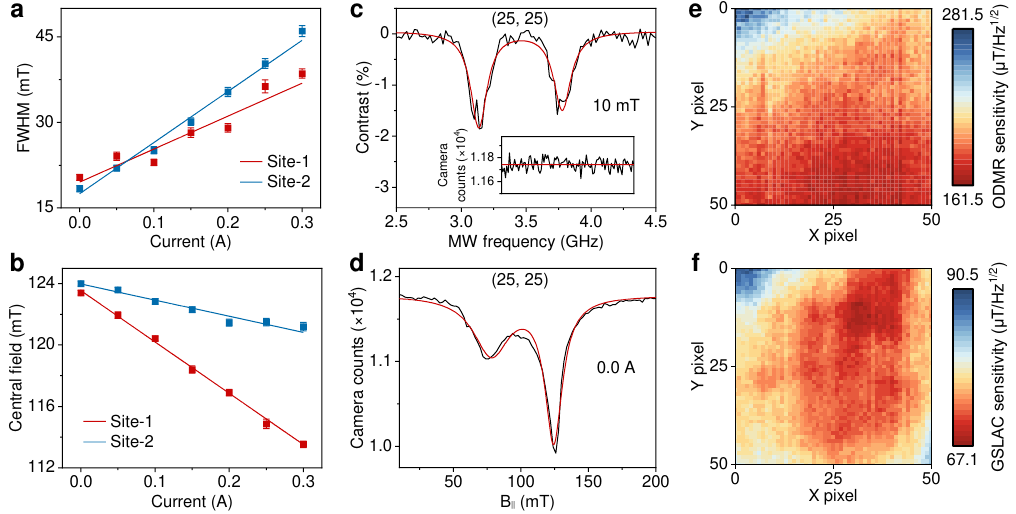}
  \caption{(a-b) FWHM (a) and central field (b) behaviors of site-1 and site-2 as a function of current magnitude. (c) ODMR spectrum under a bias magnetic field of 10 mT at (25, 25) pixel. The inset shows the referenced PL intensity of $V_B^-$ centers. (d) GSLAC spectrum without DC current at (25, 25) pixel. (e-f) Distributions of ODMR sensitivities (e) and GSLAC sensitivities (f) in an area of 10$\times$10 $\mu$m$^2$.}
  \label{fig4}
\end{figure*}

We also systematically study the relationships among central field, FWHM, and current magnitude to quantify the performance of our GSLAC scheme, by varying the current magnitude from 0.0 A to +0.3 A and conducting the corresponding measurements. The results at site-1 and site-2 pixels are shown in Fig.~\ref{fig4}(a-b). With increasing current magnitude, the FWHMs at both site-1 and site-2 exhibit a linear increase, which benefits the subsequent quantification of the $B_\perp$ component. Consistent with expectations, the FWHM at site-1 is larger than that at site-2. In Fig.~\ref{fig4}(b), with increasing current magnitude, the central fields at both site-1 and site-2 also exhibit a linear shift, which is consistent with Biot-Savart law. Therefore, both $B_\parallel$ and $B_\perp$ components can be quantitatively estimated, confirming the robustness of our method.

We further evaluate the spatial resolution and sensitivity of our setup. Given that our scheme utilizes all-optical detection to realize the magnetometry, the spatial resolution is fundamentally dominated by the optical diffraction limit and the thickness of $h$-BN film. We first calculate the optical diffraction limit of our setup. Here the numerical aperture (NA) of the objective we used is 0.75. According to Fig.~\ref{fig1}(e), the central PL wavelength $\lambda_{\text{PL}}$ of $V_B^-$ centers in our sample is about 815 nm. The optical diffraction limit is calculated by $\lambda_{\text{PL}} / (2 \cdot \text{NA}) \approx$ 0.541 $\mu$m per pixel. However, the thickness of the $h$-BN flake used in this study is measured to be approximately 0.82 $\mu$m. Accounting for additional imaging uncertainties, here the spatial resolution of our setup is conservatively estimated to be 1 $\mu$m per pixel. To comprehensively estimate the photon shot-noise-limited sensitivity of our method, we conduct both wide-field GSLAC and ODMR measurements over another 10$\times$10 $\mu$m$^2$ area under the same conditions. Referring to the calculation of photon shot-noise-limited ODMR sensitivity\cite{sensitivity1, sensitivity2}, the magnetic field sensitivity of our GSLAC scheme is given by
\begin{equation}
\eta_B = \frac{4}{3\sqrt{3}} \cdot \frac{\Gamma}{C \cdot \sqrt{R}},
\end{equation}
\begin{equation}
R = \frac{I_0 \cdot G}{Q \cdot t_{exp}},
\end{equation}
where $\Gamma$ and $C$ represent the FWHM and contrast of GSLAC dip, $R$ denotes the photon rates of $V_B^-$ centers, $I_0$ is the averaged camera raw count, $G\approx0.35$ is the gain value determined by the Full Well Capacity and analog-to-digital resolution of our camera, $Q\approx30\%$ is the quantum efficiency of the camera at 815 nm, and $t_{exp}$ is the exposure time of camera. We focus on the (25, 25) pixel of this area and estimate its ODMR and GSLAC sensitivity, respectively. Fig.~\ref{fig4}(c) exhibits its ODMR spectrum under a bias magnetic field of 10 mT. The $I_0$ is measured to be 11742 (see inset) for a $t_{exp}$ of 10 ms. The ODMR sensitivity at (25, 25) pixel is estimated to be approximately 191.3 $\mu$T/$\sqrt{\text{Hz}}$. Fig.~\ref{fig4}(d) presents the GSLAC features without DC current. The $I_0$ in Fig.~\ref{fig4}(d) is about 11787, and the corresponding sensitivity at (25, 25) pixel is estimated to be around 72.2 $\mu$T/$\sqrt{\text{Hz}}$, which represents an improvement of nearly threefold compared to the ODMR method. Furthermore, we perform sensitivity mappings for both ODMR and GSLAC schemes across this area, as shown in Fig.~\ref{fig4}(e-f). Within this region, the best ODMR sensitivity of approximately 161.5 $\mu$T/$\sqrt{\text{Hz}}$ for a single pixel, and the best GSLAC sensitivity is about 67.1 $\mu$T/$\sqrt{\text{Hz}}$ for a single pixel, also representing a near threefold enhancement.

\section{Discussion and conclusion}\label{sec4}
We develop and demonstrate an MW-free wide-field magnetometry strategy with $V_B^-$ centers in $h$-BN. Our strategy utilizes the magnetic-sensitive GSLAC mechanism of $V_B^-$ centers to optically detect the variation of external magnetic field. Combining the spin-based wide-field microscopy technique, we realize the wide-field microscopy of the induced magnetic field from a current-carrying CPW structure. We also estimate the magnetic field sensitivity and spatial resolution of our method using photon shot noise limit and optical diffraction limit, respectively. Our work extends the practical scope of $V_B^-$ centers and reduces the hardware threshold. Moreover, our technique is more suitable for the fundamental studies of microscopic magnetic sensing under extreme conditions, such as cryogenic and high-pressure scenarios\cite{highpressure1, highpressure2, highpressure3, highpressure4, highpressure5}.

Several improvement schemes can be considered further, such as increasing the quality of $V_B^-$ ensemble, choosing a high-sensitivity camera, and using advanced pulse sequences. For example, the $V_B^-$ ensemble with higher quality provides a better coherence time\cite{isotopic1, isotopic2}, which can significantly shorten the FWHM and enhance the contrast of GSLAC feature. Using lock-in camera can bring a high-speed response time and enhance sensitivity\cite{lockin}. Combining advanced pulse sequences such as double-quantum sequences\cite{dq1, dq2} and Fourier imaging sequences\cite{fourier1, fourier2} can improve the sensitivity and spatial resolution.

\subsection*{Acknowledgements}
This work was supported by the National Key Research and Development Program of China (Grant No. 2023YFF0718400), the Zhejiang Provincial Natural Science Foundation of China (Grants No. LQN26A040004 and LD25A040001), the National Natural Science Foundation of China (Grant No. 12475042), and the "Pioneer" and "Leading Goose" R\&D Program of Zhejiang Province (Grant No. 2025C01041).

\bibliography{sn-bibliography}

\end{document}